\documentclass[twocolumn,showpacs,superscriptaddress]{revtex4}

\usepackage{graphicx}\usepackage{bm}
\begin{document}
\title{
Searching for high magnetization density in bulk Fe: 
the new metastable Fe$_6$ phase
}

\author{Koichiro Umemoto} 
\affiliation{Department of Earth Sciences, University of Minnesota, 
Minneapolis, MN 55455, USA}
\affiliation{Earth-Life Science Institute, Tokyo Institute of Technology, 2-12-1-IE-1 O Okayama, Meguroku-ku, Tokyo, 152-8550, Japan}

\author{Burak Himmetoglu}
\affiliation{Materials Department, University of California,
Santa Barbara, CA 93106, USA}

\author{Jian-Ping Wang}
\affiliation{Department of Electrical and Computer Engineering, 
University of Minnesota, 
Minneapolis, MN 55455, USA}

\author{Renata M. Wentzcovitch}
\affiliation{Department of Chemical Engineering and Materials Science, University of Minneapolis, 
Minnesota 55455, USA}

\author{Matteo Cococcioni \footnote{present address: Institute of Materials, 
\'Ecole Polytechnique F\'ed\'erale de Lausanne (EPFL), 
CH-1015 Lausanne, Switzerland}}
\affiliation{Department of Chemical Engineering and Materials Science, University of Minneapolis, 
Minnesota 55455, USA}

\date{\today}
\begin{abstract}
We report the discovery of a new allotrope of iron by first principles 
calculations. This phase has $Pmn2_1$ symmetry, a six-atom 
unit cell (hence the name Fe$_6$), 
and the highest magnetization density (M$_s$) among all known 
crystalline phases of iron. 
Obtained from the structural 
optimizations of the Fe$_3$C-cementite crystal upon carbon removal,
$Pmn2_1$ Fe$_6$ is shown to result from the stabilization of a ferromagnetic
FCC phase, further strained along the Bain path.
Although metastable from 0 to 50 GPa, 
the new phase is more stable, at low pressures, than the other well-known HCP 
and FCC allotropes and smoothly transforms into the FCC phase 
under compression. 
If stabilized to room temperature, e.g., by interstitial 
impurities, Fe$_{6}$ could become the basis material for
high M$_s$ rare-earth-free permament magnets and high-impact 
applications such as, light-weight
electric engine rotors or high-density recording media.
The new phase could also be key to explain the enigmatic high M$_s$ 
of Fe$_{16}$N$_2$, which is currently attracting an intense 
research activity.  
\end{abstract}

\maketitle

Permanent magnets are at the core of applications of unquestionable
technological relevance. 
A high magnetization density (M$_s$) is certainly a very desirable 
feature in these systems as it would help the
miniaturization of magnetic devices (e.g., for high storage density
memories) and allow for lighter, more cost-effective applications.
Most of currently available permanent magnets are rare-earth (RE) compounds.
The extreme scarcity of these elements and their uneven distribution have 
stimulated, in the last decades, a vigorous search 
for magnetic materials based on earth-abundant 
elements \cite{bader14,jesche14,khan14}, sometimes with a strong 
focus on nano-structured materials \cite{nummy11,sellmyer13}).
Transition metals (TM) are the natural best alternative
to RE for fabricating magnetic devices.
On the other hand, $p$-type magnetism seems less promising 
for the above-mentioned applications.
Although elements with open $p$ shells are, in fact,
very abundant, and tend to form light-weight compounds,
the magnetization of these materials is often due to impurities and
other kinds of point defects which implies that the resulting
magnetic moments are highly diluted and lead to lower values of M$_s$.
\cite{elfimov02,droghetti08,volnianska10}.
TM-based materials offer, in this sense, better opportunities
to fabricate permanent magnets:
due to their marked atomic character and scarce hybridization,
$d$ shells remain partially filled (open) more frequently 
than $p$ shells and their higher degree of localization
makes Hund's exchange interaction stronger than for $p$ states.
For high M$_s$ applications,
3$d$ TM elements are obviously most appealing. While permanent magnetism
is possible in their compounds (in fact, it was discovered in Fe$_3$O$_4$
magnetite) the presence of other, non-magnetic elements 
obviously decreases 
their volume-specific magnetization.
In contrast, bulk elemental transition metals are usually soft magnets.
Furthermore, the partially itinerant character of their $d$ 
electrons undermines Hund's magnetic interactions and contributes 
to suppress their bulk magnetization.
Nevertheless, there has been a considerable effort to stabilize 
useful magnetic phases of transition metals through alloying 
(e.g., Fe - Ni and Fe - Co compounds) or through doping.

Within the latter scenario, Fe nitrides have attracted great interest 
\cite{cottenier09,Ji11,leinewebera01,lindmaa13,huang14}
One of the most intriguing 
Fe-N compounds, Fe$_{16}$N$_2$, 
has been the object of 
a sixty years-long debate 
about the magnetization of its $\alpha''$ ordered phase.
The commonly accepted unit cell of this material is a tetragonally 
distorted 2x2x2 supercell of bcc iron with nitrogen impurities occupying 
one fourth of the octahedral interstitial sites.
After the crystal structure of this phase 
was resolved from studying 
tempered N-doped martensites \cite{Jack1951},
most of the research on this material focused on the deposition
of high M$_s$ thin films, mostly through sputtering.
A long controversy ignited on the magnetic properties
of these films. Some groups reported
evidence of a saturation magnetization (between 2.4 and 3.2 T)
higher than that of $\alpha$ Fe (2.3 T) 
and exceeding the prediction of the Slater-Pauling theory
\cite{Kim1972,Komuro1990,Sugita1991,Okamoto1996}.
Others failed to observe any deviation from the trend
established by this theory based on itinerant magnetism
\cite{slaterp,pauling38,bozorth50}, 
also obeyed by other Fe nitrades
\cite{Ortiz1994,Takahashi1994,Sun1996,Brewer1996,Takahashi1996}. 
Recently, the interest on this material was re-sparkled, by 
research in J.-P. Wang's group. This group was able to deposit
a thin film of Fe nitride with a Fe-N stoichiometry close to 8:1 that,
after proper annealing, reproducibly showed
a magnetization density exceeding 2.68 T \cite{Ji11,Ji2011,wang13,ji13}. 

All ab-initio calculations performed in the last decade on this material,
based on the crystal structure proposed by 
Jack \cite{Jack1951}, and have failed to find 
higher M$_s$ than bulk Fe, even when using
corrections to the DFT functionals (as DFT+U, or
hybrid functionals) able to improve the description of 
electronic localization on $d$ 
states \cite{Ishida1992,Min1992,Coehoorn1993,
Coey1994,xie00,Li2006,Sims2012,ke13}. 
The only exception remains the model proposed by Wang's group \cite{Ji2010}, 
based on the existence of Fe$_6$N super-magnetic clusters with additional
Fe atoms in their interstitials. 
%
\begin{figure}[h!]
\hbox to \hsize{\hfill
\includegraphics[width=6.5cm, angle=270]{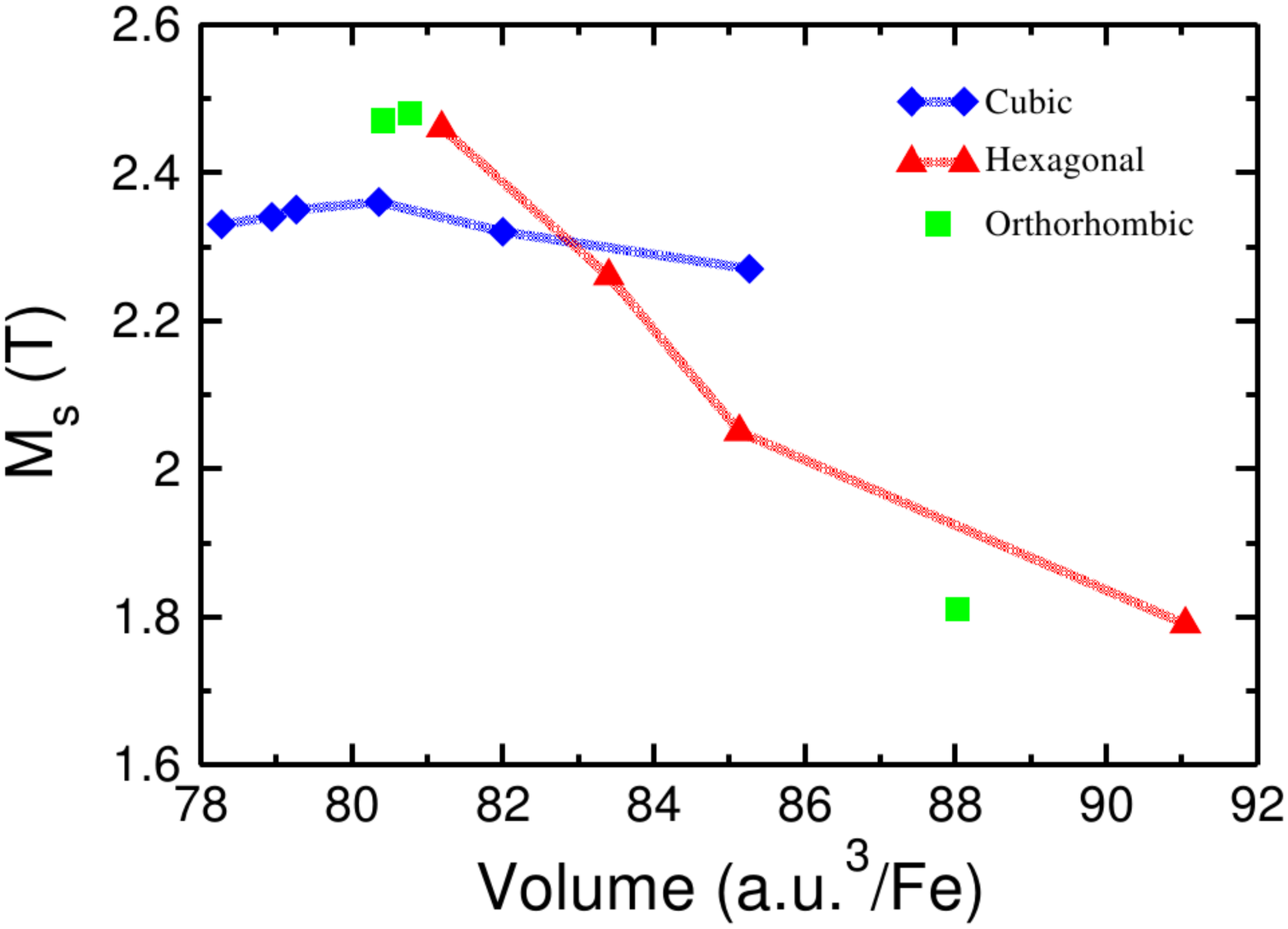}
    \hfill
 }
\caption{
(Color online)
Magnetization density (in Tesla) of several Fe-N compounds at zero
pressure. 
Lines are a guide to the eye.
From left to right blue diamonds represent, 
bcc Fe, Fe$_{64}$N, Fe$_{54}$N, Fe$_{32}$N, Fe$_{16}$N,
and Fe$_{16}$N$_2$. Red triangles represent hcp Fe, Fe$_{12}$N,
Fe$_6$N, Fe$_3$N. 
The green squares 
are relative to orthorhombic crystals and represent
$Pmn2_1$ Fe$_6$, $Pmmn$ Fe$_6$, and $Pnma$ 
Fe$_3$N cementite (see text).
}
\label{msphases}
\end{figure}

The present work is motivated by the controversial 
magnetic properties of this system.
However, in this paper we deviate from the common approach
of other works in literature and 
question the ability of the tetragonal $\alpha''$ of Fe$_{16}$N$_2$
to express a high magnetization density.
While Wang's experiments 
\cite{Ji11,Ji2011,wang13,ji13} 
suggest that the ordering of nitrogen impurities (achieved through annealing)
probably plays a key role in determining the magnetic properties 
of the obtained crystal,
the rich variety of Fe-N stoichiometries and the 
abundance of alternative crystalline phases
open other possibilities that are worth considering. 
The idea we started exploring with the present paper is to obtain
a crystal with the 16:2 stoichiometry from 
superstructures of other suitable Fe nitrides, 
whose nitrogen concentration is then properly adjusted.
Following this line, we discovered a new (metastable) allotrope of iron,
which shows the highest M$_s$ of all known phases. This new
crystal of iron could represent a first step towards the 
rationalization of the high M$_s$ of the $\alpha$'' phase of Fe$_{16}$N$_2$
and is the main focus of the present paper.

Results presented here are based on 
density-functional-theory (DFT) \cite{Hohenberg1964,Kohn1965}.
The technical details of these calculations are briefly summarized in 
the footnote \footnote{The calculations presented in this article were
performed using the plane-wave - pseudopotential codes contained
in the Quantum-ESPRESSO package \cite{Giannozzi2009}.
The exchange-correlation functional was approximated using
a generalized-gradient approximation (GGA), 
according to
the prescription of Perdew-Burke-Ernzerhof (PBE) \cite{PBE}.
The iron pseudopotential was constructed with $3s$ and $3p$ states 
in the valence
manifold and was generated by D. Vanderbilt's method \cite{Vanderbilt1990}  
using the $3s^2 3p^6 3d^{6.5} 4s^1 4p^0$ reference electronic configuration.
The expansion of electronic wavefunctions and charge density on the
plane-wave basis set was converged at kinetic energy cut-offs of
40 and 160 Ry, respectively.
A Monkhorst-Pack \cite{Monkhorst1976} $8\times 12\times 4$ 
special k-point grid was employed to integrate the Brillouin zone.
These parameters guaranteed the convergence of the total 
energy of the system to within 1 mRy/atom from its limit value.
Structural optimizations under finite pressure conditions were 
performed using the variable-cell-shape damped molecular dynamics 
technique defined in Refs. \cite{Wentzcovitch1991,Wentzcovitch1993}.
Finite temperature effects were also taken into account using the
quasi-harmonic approximation (QHA) \cite{Wallace1972}.
The vibrational spectrum of the material 
was computed by 
density-functional-perturbation theory (DFPT) \cite{Giannozzi1991,Gonze1995,
Baroni2001} as implemented in the PHONON code of the
Quantum-ESPRESSO distribution.}

\begin{figure}[h!]
\hbox to \hsize{\hfill
\includegraphics[width=8.8cm]{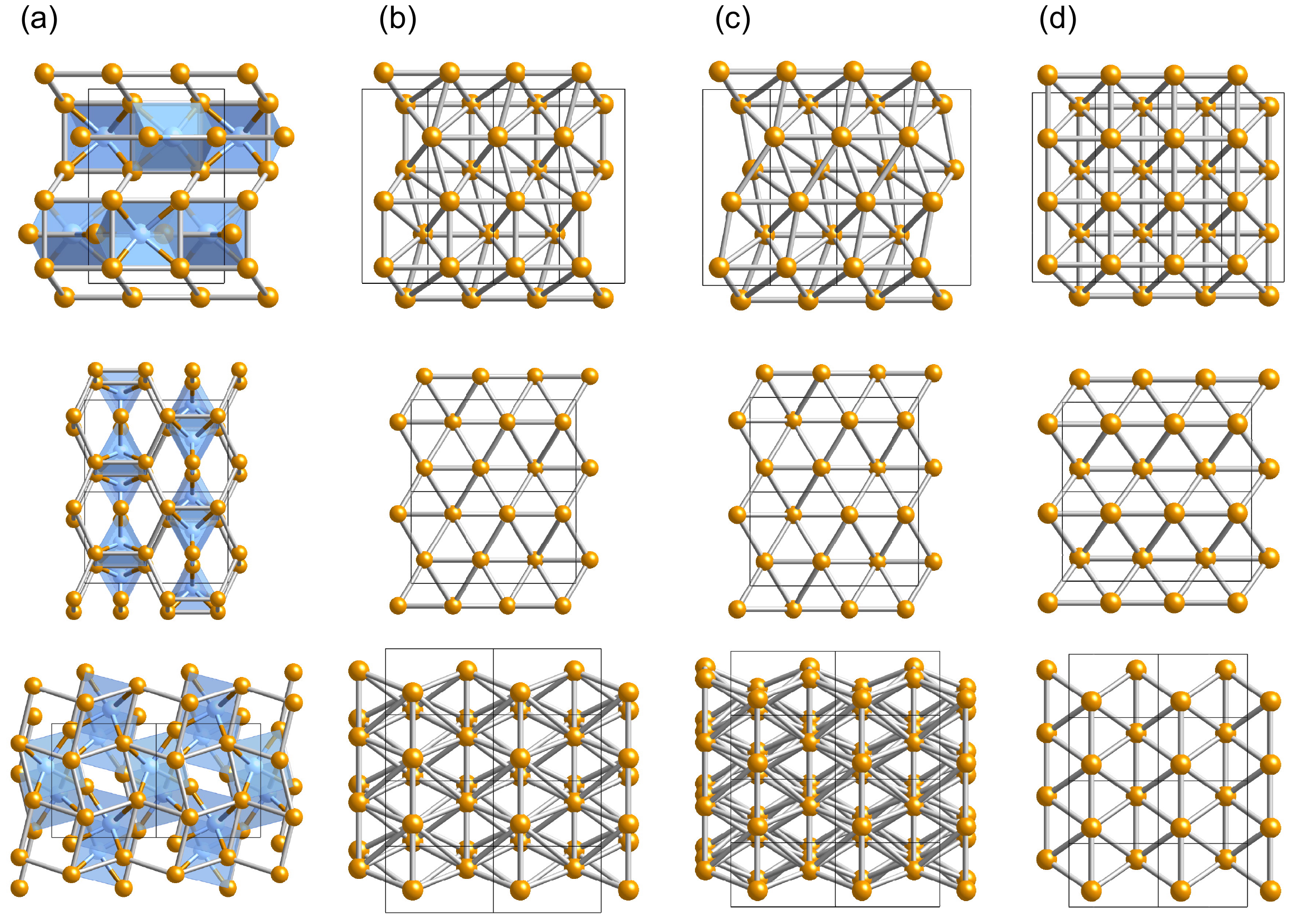}
    \hfill
 }
\caption{
Crystal structures of (a) cementite Fe$_3$N, (b) $Pmmn$ Fe$_6$, 
(c) $Pmn2_1$ Fe$_6$ at 0 GPa and (d) $Pmn2_1$ Fe$_6$ at 70 GPa 
which is identical to fcc Fe. Orange and blue spheres denote iron and 
nitrogen, respectively. Upper, middle and lower rows correspond to
the [001], [100], and [010] views of the Cementite crystal in 
column (a), to the [010], [001], and [100] views of the $Pmmn$ Fe$_6$
crystal in column (b), and to the [100], [010], and [001] views of
the $Pmn2_1$ Fe$_6$ crystals in column (c) and (d).
}
\label{structure}
\end{figure}

Our investigation started from a preliminary screening of the
saturation magnetization density of different Fe nitrides, 
whose results are shown in Fig. \ref{msphases}. 
Based on the possibility of reaching higher values of M$_s$ within the 
$\epsilon$-Fe$_3$N family of compounds (red line), this stoichiometry
was adopted as the starting point. 
However, due to the scarce resemblance with the tetragonal
structure of the $\alpha''$ phase, the hexagonal Fe$_3$N was disregarded.
On the other hand, the tendency of C-doped Fe to form,
upon slow cooling from the $\gamma$ fcc solid solution, 
an orthorhombic crystal with the same C content (Fe$_3$C cementite)
suggested the use of this structure as a starting point.
Fe$_3$C cementite has a unit cell containing four formula
units (shown in Fig.~\ref{structure}(a)) and a $Pnma$ space group. 
Its structure and symmetry 
are preserved when carbon is replaced by nitrogen, which results in a marginal
volume increase. 
Although the unit cell of
the cementite structure (Fe$_{12}$N$_4$) is not easily matchable with the
8:1 stoichiometry, it was used to qualitatively study the
effect of nitrogen removal on the magnetic and structural properties of the
crystal. 
In $Pnma$ Fe$_3$N 
nitrogen atoms occupy the center of trigonal prisms of irons, 
and are six-fold coordinated.
The magnetic moment (per atom) 
and magnetization density of Fe$_3$N are relatively low:
2.05 $\mu_B$/Fe and 1.81 T, respectively. 
The removal of one nitrogen per cell 
does not affect significantly the structure of the crystal. 
Instead, when two or more nitrogen atoms are subsequently
removed from the unit cell,
giving Fe$_6$N, Fe$_{12}$N and pure Fe, respectively, structural
optimization leads to a drastic change of the crystal structure,
We focused our attention on the pure iron phase which assumes Pmmn symmetry.
The $Pmmn$ crystal structure, obtained upon extraction of all
nitrogens from $Pnma$-type Fe$_3$N, is shown in Fig.~\ref{structure}(b).
Due to a translational symmetry, not present in the original 
$Pnma$ crystal, the unit cell of the $Pmmn$ Fe is half the size 
of that of cementite 
with six atoms instead of twelve
(hence the name Fe$_6$).
The rest of this paper will be dedicated
to discussing the properties of the Fe$_6$ crystal. 

\begin{figure}
\hbox to \hsize{\hfill
\includegraphics[width=9cm]{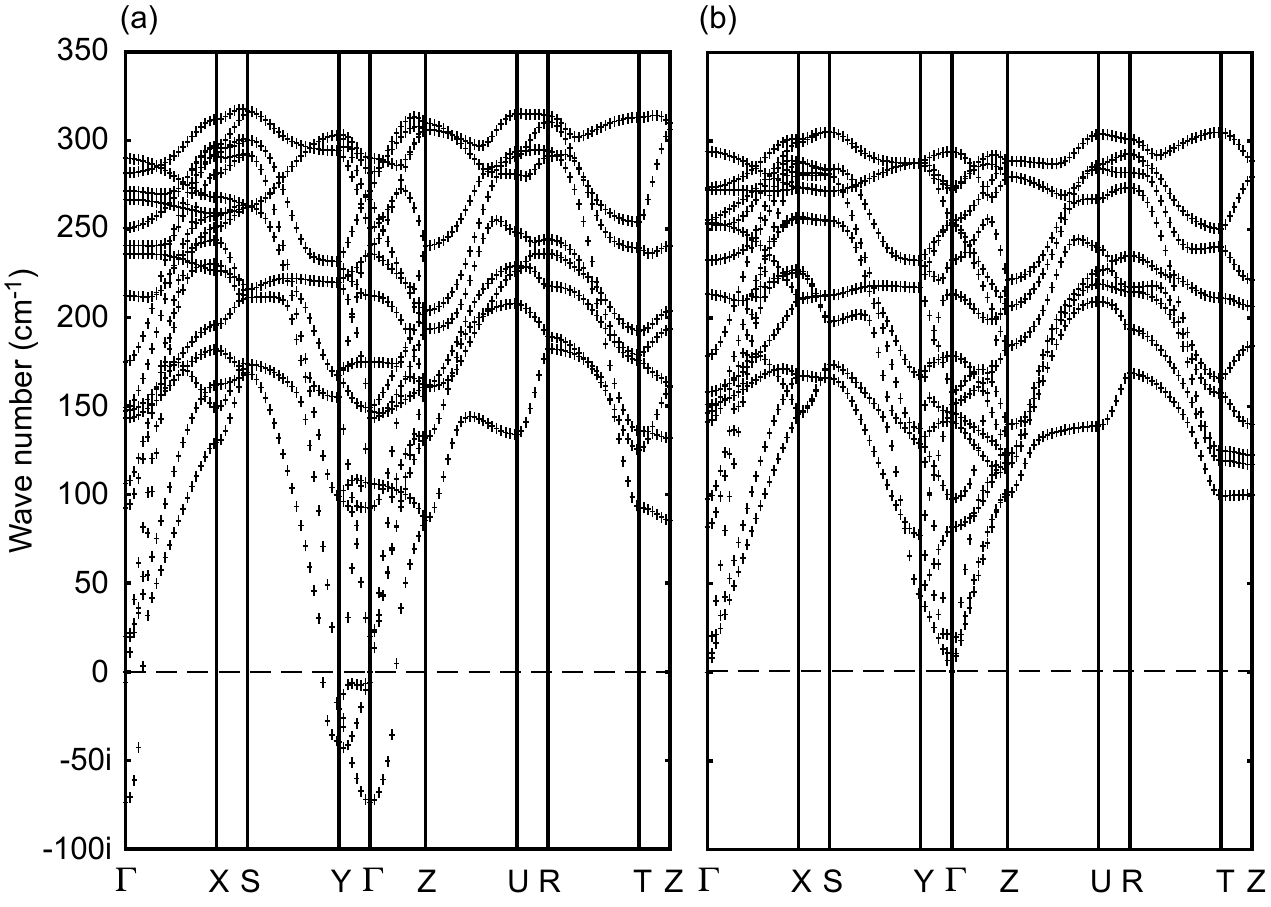}
    \hfill
 }
\caption{
Phonon dispersions at 0 GPa of (a) interim $Pmmn$ Fe$_6$ and (b) $Pmn2_1$ Fe$_6$.
}
\label{phonon}
\end{figure}

The $Pmmn$ crystal was obtained by a naturally symmetry-preserving
structural optimization \cite{rmw91}. Its phonon dispersions, reported in 
Fig.~\ref{phonon}(a), still indicate the presence of unstable modes
in the vicinity of $\Gamma$ and $Y$ points in the Brillouin zones.
Structural re-optimizations along the pattern of atomic displacements
of the unstable zone-center mode produces a new structure with
symmetry $Pmn2_1$, shown in Fig.~\ref{structure}(c).
This structure is dynamically stable as proved by the absence of 
imaginary frequencies in its vibrational spectrum 
(see Fig.~\ref{phonon}(b)).
To the best of our knowledge this 
allotrope of iron has not been reported in literature.

The new phase crystal structure is
shown in column (c) of Fig.~\ref{structure}.
As evident from the top figure, it presents 
a three-fold modulation on the (001) planes
with a asymmetric (two-to-one)
``zig-zag" alternation of distorted Fe quadrangles.
If one orientation of distorted quadrangles is prevalent 
on one (001) plane, the opposite occurs on neighbor 
parallel planes.  
The resulting ABAB stacking along [001] prevents the formation of
mirror planes in the crystal.
A comparison with $Pmmn$ Fe$_6$ (column (b)
of Fig.~\ref{structure}) highlights the relationship among the two
structures and the deformation induced by the optimization
of the $Pmmn$ crystal along its $\Gamma$ soft mode. 
The space group of the new optimized structure, $Pmn2_1$, is a subgroup 
of $Pmmn$. Hereafter, we refer to this new phase as ``$Pmn2_1$ Fe$_6$''.
Lattice parameters and atomic positions for the $Pmmn$ and $Pmn2_1$
Fe$_6$ crystals are reported and compared in Table \ref{data} using the Wyckoff 
notation. 
%
%
\begin{table}
\begin{tabular}{ccc}
\hline
\hline
\multicolumn{3}{c}{$Pmn2_1$-type Fe$_6$} \\
\hline
$(a,b,c)$ (\AA) & & (4.032, 2.464, 7.197) \\
\hline
Fe$_1$ & $(2a)$ & $(0, 0.07044, 0.75564)$ \\
Fe$_2$ & $(2a)$ & $(0, 0.25046, 0.08932) $ \\
Fe$_3$ & $(2a)$ & $(0, 0.42873, 0.42311) $ \\
\hline
\hline
\multicolumn{3}{c}{$Pmmn$-type Fe$_6$} \\
\hline
$(a,b,c)$ (\AA) & & (7.252, 4.026, 2.460) \\
\hline
Fe$_1$ & $(2b)$ & $(0, 1/2, 0.93204)$ \\
Fe$_2$ & $(4f)$ & $(0.83069, 0, 0.33549) $ \\
\hline
\end{tabular}
\caption{
Calculated lattice constants and atomic Wyckoff coordinates $x, y, z$ of
$Pmn2_1$ and $Pmmn$ Fe$_6$ at 0 GPa.
The $(2a)$, $(2b)$, and $(4f)$ positions are given by $\{(0,y,z), (1/2,-y,z+1/2)\}$,
$\{(0,1/2,z), (1/2,0,-z)\}$, and $\{(x,0,z), (-x,0,z), (-x+1/2,1/2,-z), (x+1/2,1/2,-z)\}$, respectively.
}
\label{data}
\end{table}
%

\begin{figure}[h!]
\hbox to \hsize{\hfill

\includegraphics[width=7cm,angle=270]{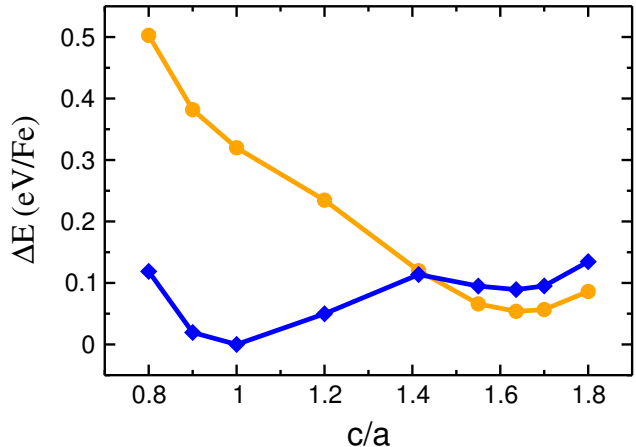}
    \hfill
 }
\caption{
(Color online)
Deformation energy of a 1$\times$3$\times$1 bcc/bct supercell (blue line)
and the $Pmn2_1$ Fe$_6$ cell (orange line). The deformations are imposed
maintaining the volume constant and equal to the equilibrium
volume of $Pmn2_1$ Fe$_6$.}
\label{enerca}
\end{figure}

Since the total energy at 0 GPa
obtained for this phase is 4.2 mRy/Fe higher than 
that of the ferromagnetic (FM) bcc crystal, $Pmn2_1$ Fe$_6$ is a metastable
phase of iron.
Its M$_s$ (2.47 T) is the highest ever reported for any (meta)stable
allotrope of this elemental metal.
The unit cell of $Pmn2_1$ Fe$_6$ can be understood as
derived from a 1$\times$3$\times$1 supercell of a conventional bcc
structure (simple cubic lattice with a two-atoms basis) with a slight
contraction along [010] ($b/a\sim 2.92$), a substantial 
elongation along [001] ($c/a\sim 1.65$) and a significant readjustment of
ionic positions, producing the above-mentioned structural modulation.
Indeed, the appearance of metastable modulated phases along the Bain 
deformation path from 
the high-temperature austenite to the low-temperature 
martensite phases is not uncommon in materials undergoing martensitic
transformations. 
In $Pmn2_1$ Fe$_6$, however, the $c/a$ ratio exceeds that of fcc
($c/a = \sqrt{2}$) thus preventing its classification as an intermediate
phase between bcc ($c/a = 1$) and fcc.
Figure \ref{enerca} 
shows volume-preserving deformation energies versus $c/a$
of (a slightly deformed) 
$Pmn2_1$ Fe$_6$ (orange line) and of a $1\times 3\times 1$
supercell of conventional bcc/bct structure (blue line).
%
The volume, equal to that of the equilibrium structure of
$Pmn2_1$ Fe$_6$, is 3.8\% larger than that of the bcc structure
at equilibrium, and 2.3\% smaller than that of the fcc crystal optimized
with a FM configuration.
As seen from the figure, the tetragonal deformation
of the bcc cell results in a double-well energy profile in which the bcc
structure ($c/a = 1$) represents the global minimum, the fcc 
($c/a = \sqrt{2}$) the central maximum, a bct cell with $c/a\sim 1.64$
(compatible with that of $Pmn2_1$ Fe$_6$) the second (local) minimum. 
The orange curve in the figure is obtained by
changing the value of $c/a$ in an orthorhombic cell with atomic positions
compatible to those of the modulated $Pmn2_1$ Fe$_6$
structure  and $b/a = 3$, instead of the equilibrium value 2.92, 
for a better comparison with the bcc supercell.
This curve has only one minimum, in correspondence of the Fe$_6$
cell, which is deeper than the local minimum of
the distorted bcc supercell energy profile. 
The two energy curves cross-over
in correspondance of the fcc structure, which thus plays the
role of a saddle point between the bcc and the $Pmn2_1$ Fe$_6$
basins, at zero pressure. 

\begin{figure}[h!]
\hbox to \hsize{\hfill

\includegraphics[width=7cm,angle=270]{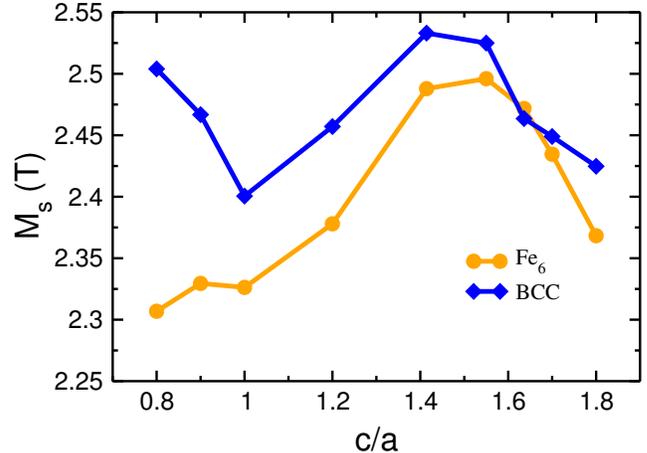}
    \hfill
 }
\caption{
(Color online)
The magnetization density as a function of $c/a$ for a distorted 
1$\times$3$\times$1 bcc supercell (blue line) and for the orthorhombic 
$Pmn2_1$ Fe$_6$ (orange line). 
}
\label{msca}
\end{figure}

Being obtained from a ferromagnetic crystal,
$Pmn2_1$ Fe$_6$ also has a FM ground state.
Its AFM and non-magnetic (NM) phases (the latter is actually
coincident with fcc, as discussed below) have energies
36 and 75 mRy/atom higher than that of the FM phase, 
respectively. 
The magnetic moment and magnetization density of this phase are, respectively,
2.55 $\mu_B$/Fe and 2.47 T, quite significantly higher than those of the bcc 
phase (2.34 $\mu_B$/Fe or 2.33 T). In fact, to the best of our
knowledge, these values are the highest
ever reported for any allotrope of Fe. 
In order to understand the origin of the high magnetization density
it is useful to study how this quantity 
changes with $c/a$ for the two types of distorted cells
considered above. Results, 
shown in Fig. \ref{msca},
indicate that the bcc supercell has a higher magnetization
density than $Pmn2_1$ Fe$_6$ for all the considered values of 
$c/a$, except those in
the close vicinity of the Fe$_6$ stability basin, where the two curves 
essentially coincide. This observation clarifies that the modulation in
the $Pmn2_1$ Fe$_6$ structure, i.e., the shift of atoms from 
their bcc equilibrium positions, is only responsible for the stabilization of
the Fe$_6$ orthorhombic crystal 
but plays a very marginal role in the 
on-set of its high magnetization. 
The deviation of $b/a$ from 3 also has negligible effects on M$_s$.
The high magnetization density is, instead,
mostly due to the elongation of the $c$ axis 
that transforms the bcc into
a bct cell with $c/a\sim 1.64$. 
This distortion of the bcc structure
changes the order (in energy) of the 3$d$ states of iron and allows
their occupancies to conform to Hund's rules more closely, 
thus enhancing magnetization.
It is important to note that the FM fcc structure 
has the highest M$_s (\sim2.53$ T) among all
the phases considered in this study; at the same time, the 
bcc structure ($c/a = 1$) corresponds to the minimum of M$_s$
along the bcc curve. Because of deviations from
their equilibrium volumes, the ground state values of M$_s$ for bcc and
FM fcc are different from those reported in Fig. \ref{msphases} and
result in 2.33 T and 2.56 T, respectively. 
%

\begin{figure}
\hbox to \hsize{\hfill
\includegraphics[width=8.5cm]{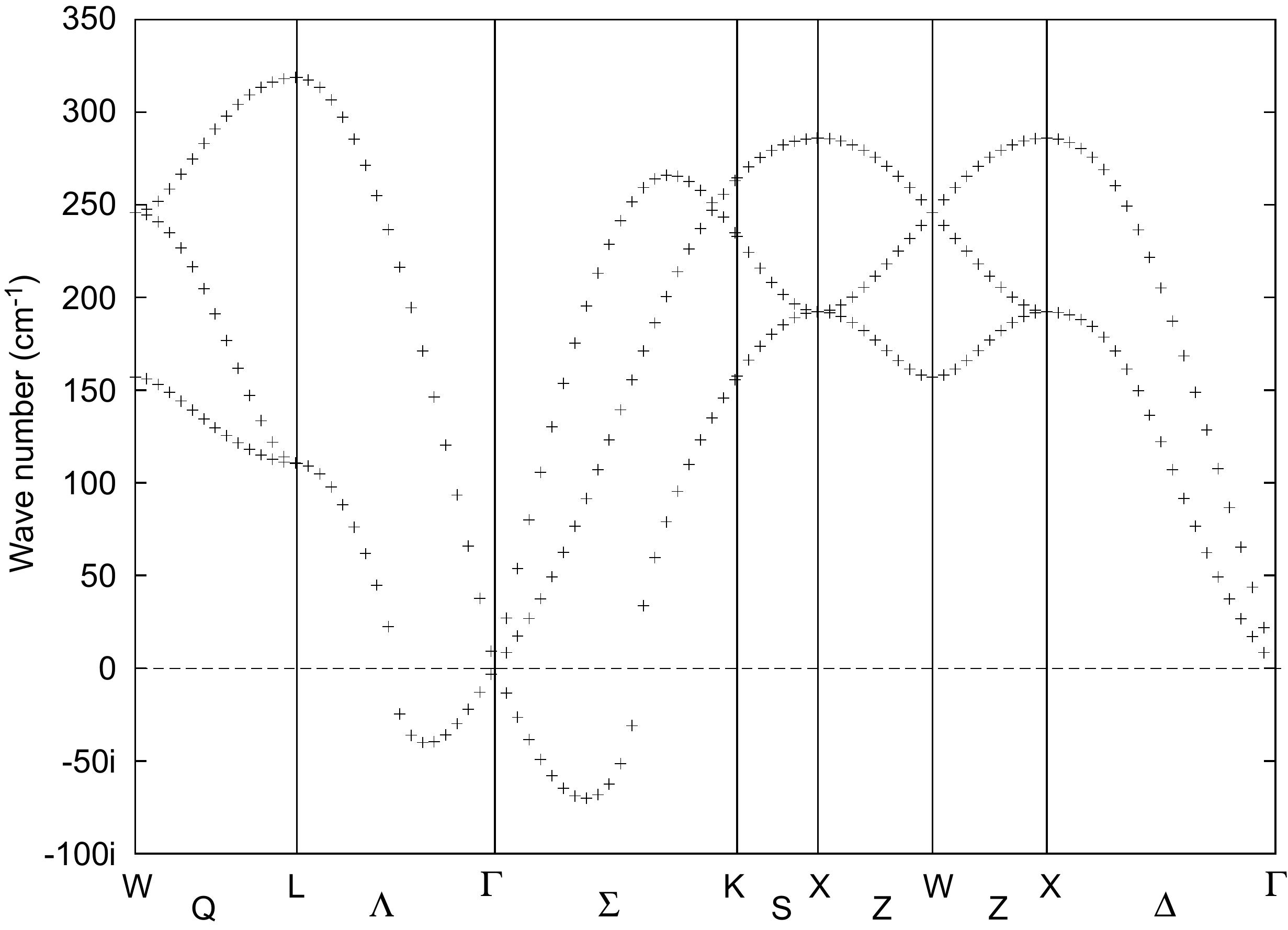}
    \hfill
 }
\caption{Phonon dispersion of FM fcc iron at 0 GPa.}
\label{fcc}
\end{figure}

The magnetic ground state of fcc Fe has been reported to 
be non-collinear and to consist of a spin spiral 
incommensurate with the lattice \cite{Tsunoda1993,Uhl1994}. 
The spin modulation of the fcc phase represents
an intriguing counterpart of 
the structural modulation that leads to $Pmn2_1$ Fe$_6$ and
stabilizes a FM ground state. 
To further investigate this aspect, we decided to analyze
the vibrational spectrum of the FM fcc crystal. 
The obtained phonon dispersions indicate 
the presence of two soft modes
along the $\Sigma$ and $\Lambda$ lines (see Fig. \ref{fcc}).
Soft modes along the $\Lambda$ line are not very interesting
to the present study. In fact, optimizing the unit cell 
with an ionic displacement pattern corresponding, for example,
to the imaginary-frequency mode at 
$(1/6, 1/6, 1/6)$ merely evolves the fcc structure
into the bcc one, along the very well known Bain deformation path.
The soft modes along the $\Sigma$ line are, instead, more interesting.
The largest imaginary phonon frequency ($\sim$70$i$ cm$^{-1}$) 
occurs at 
$(\xi,\xi,0)$ with $\xi \sim 0.4$. 
To clarify the nature of the structural deformation induced by this soft 
branch, we studied the soft vibrational mode 
at $(1/3, 1/3, 0)$, the closest to the minimum of the soft phonon line
that is compatible with a supercell of reasonable size 
(3$\times$3$\times$1). 
By optimizing this supercell of the fcc crystal 
with atoms moved from their equilibrium positions
according to the displacement pattern of this soft mode,
a three-fold modulated array of distorted Fe octahedra is obtained. 
The FM fcc phase transforms into the interim $Pmmn$ structure, 
shown in Fig.~\ref{structure}(b) and, eventually, into $Pmn2_1$ Fe$_6$,
thus highlighting its direct connection with this allotrope. 
The position of the FM fcc crystal at a saddle point of the energy
profile between the bcc and the Fe$_6$ structures, that emerged from Fig.
\ref{enerca}, is thus fully
confirmed by the study of its vibrational properties.
The presence of a whole band of soft modes along the $\Sigma$ direction
of the Brillouin zone indicates instabilities in the FM FCC crystal
at all length scales.
%
The ability to undergo structural distortions, either commensurate,
as the one leading to $Pmn2_1$ Fe$_6$, or incommensurate,
could probably be understood as the structural counterpart of the incommensurate
magnetic structure (spin-spiral) of the fcc crystal \cite{Tsunoda1993,Uhl1994}.
In short, while a perfectly periodic fcc crystal is 
characterized by an incommensurate spin-spiral order, a FM ground
state can be only stabilized by structural modulations with various
periodicities.

\begin{figure}[h!]
\hbox to \hsize{\hfill

\includegraphics[width=8.7cm]{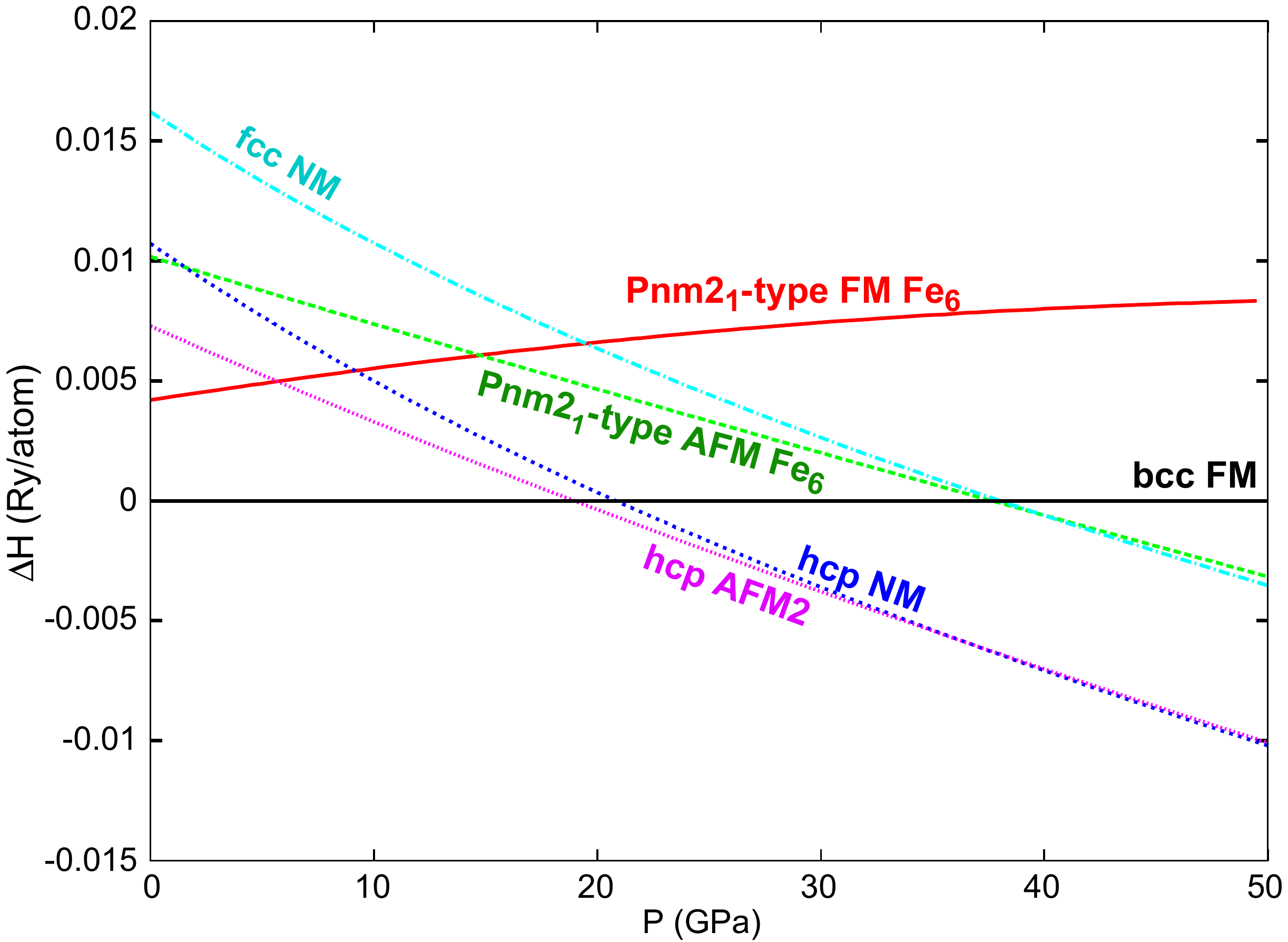}
    \hfill
 }
\caption{
Relative enthalpies of several iron allotropes with respect to bcc FM iron.}
\label{dH}
\end{figure}

It is important to keep in mind that the volume of the FM FCC phase
considered above is larger than that of its non magnetic counterpart, 
and of both bcc and Fe$_6$ phases. 
In order to clarify how their relative stability changes
with volume and to better understand the relationship between
the $Pmn2_1$ Fe$_6$ crystal and other phases of Fe, it is useful
to study the behavior of this allotrope under hydrostatic pressure.
Fig.~\ref{dH} shows the enthalpies of different crystalline structures of Fe
(with the FM bcc taken as reference) for pressures between 0 and 50 GPa.
At 0 GPa $Pmn2_1$ Fe$_6$ is more stable than all the other allotropes,
and its total energy is 2.9 mRy/atom lower than that of the AFM type-II hcp
crystal, a high-pressure form of pristine iron \cite{Steinle-Neumann2004}
and, up to now, the second most stable phase, after bcc.
Finite temperature effects, accounted for by including vibrational
terms into the free energy, e.g. within 
the quasi-harmonic approximation \cite{Wallace1972},
hardly affect the stability. The $Pmn2_1$ Fe$_6$ remains metastable 
with respect to the bcc phase at lower pressures.
Its free energy at 300 K is 3.9 mRy/atom 
higher than that of bcc Fe and,
even if quantitative details, such as transition pressures, may change 
with temperature, 
the overall qualitative picture emerging from Fig.~\ref{dH} remains 
unaltered.


Under pressure $Pmn2_1$ Fe$_6$ undergoes a magnetic transition
into an AFM phase (at $\sim$15 GPa), followed by a
continuous transformation 
into the NM fcc phase, completed at $\sim$40 GPa. 
Thus we can evince a sort of ``magneto-structural duality" between 
$Pmn2_1$ Fe$_6$ and fcc Fe:
while the FM ground state of the fcc phase can be stabilized by a
three-fold modulation compatible with the structure of the $Pmn2_1$ Fe$_6$,
$Pmn2_1$ Fe$_6$ transforms into fcc under pressure, 
upon losing its magnetization. 

In summary, we found a new metastable allotrope 
of bulk Fe that is characterized by a unit cell of six atoms with a 
$Pmn2_1$ space group, and a magnetization density higher than any other
known phase.  
Investigation of this structure highlighted its connection with the bcc
and fcc crystals and showed that it can be obtained from a 
modulation of a 1$\times$3$\times$1 bcc supercell
after imposing an elongation of the $c$ axis up to
$c/a$ values of $\sim1.64$. Along the extended Bain deformation path
that transforms the bcc structure into the $Pmn2_1$ Fe$_6$, a 
collinear-ferromagnetic fcc phase represents an intermediate  
maximum of the energy (a saddle point) between the bcc
and the $Pmn2_1$ Fe$_6$ phases, and corresponds to the largest
bain deformation that does not produce a modulation of 
atomic positions.
This picture is confirmed by the instability of fcc towards
both bcc and $Pmn2_1$ Fe$_6$.

Owing to its high magnetization density, a stabilized Fe$_6$ could
be the starting point to design materials with possibly high magnetic
coercivity, for highly technological applications as, for example,
lighter-weight electrical engine and power generator rotors, or 
high information density recording media.
In fact, it would be interesting to verify whether a high 
magnetization-density material can be obtained from $Pmn2_1$ 
Fe$_6$ by adding nitrogen impurities and, in particular, with a stoichiometry 
close to that of Fe$_{16}$N$_2$.
Unfortunately, the dimension and the 
complexity of the needed supercell makes this idea quite challenging
to test computationally.
From the abundant literature on iron and its well consolidated 
metallurgy, there seems to be little/no evidence that a FM
fcc phase can be stabilized at room temperature through addition of doping 
impurities (except, perhaps, in nanoparticles \cite{wei06}). 
However, re-interpreting the structure
of $\alpha''$ Fe$_{16}$N$_2$ as a stabilized fcc supercell could provide a key 
insight to explain its enigmatic giant M$_s$, and will be the object
of future investigations.

%
%
%

\section{\bf Acknowledgments.}
Authors are grateful to Prof. R. D. James, Xiaowei Zhang, and Dr. Yanfeng Jiang
for very useful discussions. 
They also thank Daniele Dragoni and Prof. N. Marzari for critically 
reading the manuscript. 
The work was supported by the 
US Department of Energy ARPA-E (Advanced Research Projects Agency-Energy) 
REACT program under Contract No. 0472-1595.
MC acknowledges partial support from the NSF CAREER
Award No. DMR 1151738; RMW and MC from the NSF grant No. EAR 1319361.
Calculations were performed at the Minnesota Supercomputing Institute and at 
the Laboratory for Computational Science and Engineering at the 
University of Minnesota.  

\bibliographystyle{unsrt}
\bibliography{refs}

\newpage

\end{document}